\documentclass[reprint,aps,pra,longbibliography,superscriptaddress]{revtex4-2}

\usepackage[T1]{fontenc}
\usepackage[utf8]{inputenc}
\usepackage{color}
\usepackage{babel}
\usepackage{amsmath}
\usepackage{amssymb}
\usepackage{subfigure}
\usepackage{graphicx}
\usepackage{physics} 
\usepackage{dsfont}
\usepackage{hyperref}
\usepackage{svg}
\usepackage{subcaption}
\usepackage{epsfig}
\usepackage{ragged2e}
\usepackage[normalem]{ulem}
\usepackage{comment}

\begin{document}

\title{Asymmetry and irreversibility in Lipkin-Meshkov-Glick model in the dynamical critical regime}

\author{Andesson B. Nascimento}
\email{andessonnascimento@discente.ufg.br}
\affiliation{QPequi Group, Institute of Physics, Federal University of Goi\'as, Goi\^ania, Goi\'as, 74.690-900, Brazil}

\author{Lucas C. C\'eleri\href{https://orcid.org/0000-0001-5120-8176}{\includegraphics[scale=0.05]{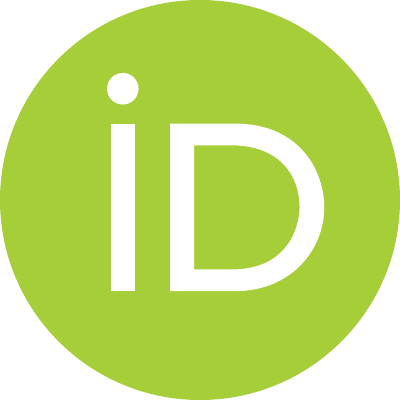}}}
\email{lucas@qpequi.com}
\affiliation{QPequi Group, Institute of Physics, Federal University of Goi\'as, Goi\^ania, Goi\'as, 74.690-900, Brazil}

\begin{abstract}
Symmetries play a central role in both equilibrium and nonequilibrium phase transitions, yet their quantitative characterization in dynamical quantum phase transitions (DQPTs) remains an open challenge. In this work, we establish a direct connection between the symmetry properties of a many-body model and measures of quantum asymmetry, showing that asymmetry monotones provide a robust and physically transparent indicator of dynamical quantum criticality. Focusing on the quenched Lipkin–Meshkov–Glick model, we demonstrate that asymmetry measures associated with collective spin generators faithfully capture the onset of DQPTs, reflecting the dynamical restoration or breaking of underlying symmetries. Remarkably, the time-averaged asymmetry exhibits clear signatures of the dynamical critical point, in close correspondence with both the dynamical order parameter and the behavior of entropy production. We further uncover a quantitative link between asymmetry generation and thermodynamic irreversibility, showing that peaks in asymmetry coincide with maximal entropy production across the transition. Our results position asymmetry as a unifying concept bridging symmetry, information-theoretic quantifiers, and nonequilibrium thermodynamics in dynamical quantum phase transitions, providing a powerful framework for understanding critical dynamics beyond traditional order parameters.
\end{abstract}

\maketitle

\section{Introduction}
\label{secI}

Symmetry principles play a central role in modern physics, providing powerful constraints on both equilibrium and nonequilibrium dynamics. In quantum systems, symmetries are traditionally associated with conserved quantities through Noether’s theorem; however, they also govern the structure of quantum coherence and the distribution of information between symmetry sectors. As a result, symmetry considerations are deeply intertwined not only with phase transitions but also with the emergence of irreversibility and entropy production in driven quantum systems~\cite{Marvian2014}. Understanding how symmetries manifest themselves dynamically is, therefore, essential for a comprehensive description of nonequilibrium quantum phenomena.

The Lipkin-Meshkov-Glick (LMG) model provides a paradigmatic setting in which these ideas can be explored~\cite{LMG1965a,LMG1965b,LMG1965c}. This fully connected spin system exhibits long-range interactions and possesses two fundamental symmetries: permutation symmetry, reflecting the collective nature of the interactions~\cite{Sciolla2011}, and a discrete $\mathbb{Z}_2$ (parity) symmetry associated with spin inversion in the $zy$ plane~\cite{Ribeiro2008,Castanos2006,Mzaouali2021}. Although both symmetries play an important role in the equilibrium quantum phase transition of the model, the $\mathbb{Z}_2$ symmetry is particularly significant, as its spontaneous breaking distinguishes the paramagnetic and ferromagnetic phases.

In recent years, the study of dynamical quantum phase transitions (DQPTs) has revealed that symmetry considerations remain equally fundamental far from equilibrium~\cite{Heyl2013,Heyl2018,Heyl2019}. DQPTs are commonly identified either by nonanalyticities in the Loschmidt echo rate function (type-II DQPTs)~\cite{Heyl2014,Heyl2015,Vajna2015,Bhattacharya2017,Heyl2017a,Flaschner2018,Jurcevic2017,Goes2020}, or by dynamical order parameters that signal symmetry restoration or breaking during post-quench evolution (type-I DQPTs)~\cite{Sciolla2013, Smacchia2015,Halimeh2017,Zhang2017,Chen2020,Bento2024}. The LMG model has served as a particularly fruitful platform in this context~\cite{Heyl2018, Muniz2020, Bento2024, Zunkovic2016}, as it allows for a clear identification of dynamical critical points and their relation to the underlying symmetries.

Despite this progress, a quantitative characterization of how symmetry is dynamically broken or restored across a DQPT --- and how this process relates to nonequilibrium thermodynamics --- remains incomplete. In particular, traditional observables, such as conserved quantities or order parameters, are often insensitive to the coherence between different symmetry sectors that develops during unitary dynamics. This motivates the search for alternative diagnostics capable of capturing symmetry breaking at the level of quantum coherence.

In this regard, the concept of quantum asymmetry has emerged as a powerful framework. Originally developed within the resource theory of quantum reference frames, asymmetry quantifies the extent to which a quantum state breaks a given symmetry, being directly associated with coherence relative to the eigenbasis of symmetry generators~\cite{Gour2008,Gour2009,Skotiniotis2012,Toloui2011,Vaccaro2008}. A comprehensive theory of asymmetry measures was developed in Refs.~\cite{Marvian2013,Marvian2014b,Marvian2016a,Marvian2016b}, where it was shown that such measures capture aspects of quantum states that are invisible to conserved quantities alone. Importantly, asymmetry measures provide a natural bridge between symmetry, coherence, and information-theoretic notions, making them particularly appealing for the study of nonequilibrium quantum dynamics.

In this work, we explore whether asymmetry measures can serve as reliable and physically meaningful indicators of dynamical quantum phase transitions. Focusing on the quenched LMG model, we analyze the time evolution of an $\ell_1$-norm–based asymmetry measure $F_L(\rho)$ with respect to collective spin generators $J_x$, $J_y$, and $J_z$. We show that the time-averaged asymmetry exhibits clear signatures of dynamical criticality, faithfully tracking the location of the DQPT and its dependence on the anisotropy parameter. Furthermore, we demonstrate a close correspondence between asymmetry, entropy production, and the dynamical order parameter, establishing asymmetry as a unifying quantity that links symmetry breaking, information-theoretic coherence, and nonequilibrium thermodynamics in dynamical phase transitions.

The paper is organized as follows. In Sec.~\ref{SecII}, we introduce the model and discuss the role of its symmetries in both equilibrium and dynamical phase transitions. In Sec.~\ref{SecIII}, we review the concept of quantum asymmetry and introduce the asymmetry measure employed in this work. Section~\ref{SecIV} presents our main results on asymmetry and dynamical criticality in the LMG model. Finally, Sec.~\ref{SecV} summarizes our conclusions and outlines possible directions for future research.

\section{The model}
\label{SecII}

Although our results apply to other models exhibiting dynamical criticality, to be concrete, we consider here the Lipkin-Meshkov-Glick model~\cite{LMG1965a,LMG1965b,LMG1965c}, which describes a system of $N$ spin-$1/2$ particles with all-to-all interactions, subjected to a transverse magnetic field. Its Hamiltonian reads
\begin{equation}
    H = -\frac{J}{j}\left( J_{z}^{2} + \gamma J_{y}^{2}\right) - 2hJ_x,
    \label{H_LMG}
\end{equation}
where $J$ is the interaction strength in the $zy$-plane, $j = N/2$ is the total angular momentum, while $h$ is the transverse magnetic field. The operators $J_l = \sum_{i=1}^{N} \sigma_l^i/2$ represent collective spin components in the $l = x, y, z$ directions, with $\sigma_l^i$ denoting the $l$ Pauli matrix acting on the $i$-th spin. The parameter $\gamma \in [0,1]$ controls the anisotropy of the interaction: $\gamma = 0$ corresponds to the maximally anisotropic case, while $\gamma = 1$ recovers the isotropic limit~\cite{Pappalardi2024}.

The LMG model possesses two important symmetries: permutation symmetry~\cite{Ribeiro2007}, which refers to invariance under particle exchange due to the equal interaction strength between all spins in the chain; and $\mathbb{Z}_2$ symmetry~\cite{Mzaouali2021,Castanos2006}, also known as parity symmetry, corresponding to invariance under transformation $J_y \to -J_y$ and $J_z \to -J_z$, while $J_x$ remains unchanged. This latter symmetry arises from the fact that the interaction terms in the Hamiltonian are quadratic in $J_y$ and $J_z$, whereas the transverse field couples linearly to $J_x$. As a result, the Hamiltonian is invariant under spin inversion in the $zy$-plane.

The $\mathbb{Z}_2$ symmetry plays a fundamental role in both the equilibrium and dynamical phase transitions of the model. For example, in the context of quantum phase transitions, the system can change from a phase in which the ground state preserves the $\mathbb{Z}_2$ symmetry, the paramagnetic phase ($h \gg J$), where the transverse field dominates and aligns the collective spin along the $x$ direction, into a symmetry-broken phase. In the ferromagnetic regime ($h \ll J$), the spin-spin interactions become dominant, and the system spontaneously selects one of two degenerate ground states that break the $\mathbb{Z}_2$ symmetry~\cite{Ribeiro2007}. A similar scenario occurs in dynamical quantum phase transitions, where the system can dynamically evolve between phases with distinct symmetry properties after a quench across the dynamical critical point~\cite{Pappalardi2024}.

The anisotropy parameter $\gamma$ controls the balance between the quadratic interaction terms. For $\gamma = 0$, the model is fully anisotropic, and the interaction occurs along a single spin component, leading the ground state at $h_0 = 0$ to align with that direction, the $z$-axis in our model. In terms of the Bloch sphere representation, we can see the ground state as a collective spin in the $z$-direction.

As $\gamma$ increases, the term $J_y^2$ gains weight, causing the ground state to gradually tilt within the $zy$-plane. When $J_z^2$ remains dominant, the ground state is still close to that of the fully anisotropic case, but this tilt becomes more pronounced as $\gamma \to 1$.

In contrast, in the isotropic case $\gamma = 1$, the Hamiltonian in Eq.~(\ref{H_LMG}) can be rewritten in terms of $J_x^2$ only, using the identity $\mathbf{J}^2 = J_x^2 + J_y^2 + J_z^2$. Therefore, for $h_0 = 0$, the ground state of $H_0$ corresponds to one of the degenerate states along the $x$-axis, a collective spin in the $x$-direction in the Bloch sphere~\cite{Vidal2004}.

Let us now consider quench dynamics in which the magnetic field $h$ suddenly changes from $h_0$ to some final value, and the system is allowed to evolve under the new Hamiltonian. In this situation, the LMG model exhibits dynamical quantum phase transitions that can be identified through two complementary approaches: the rate function associated with the return probability $\mathcal{L}(t)=|\bra{\psi_0}e^{-iHt}\ket{\psi_0}|^2$, with $\ket{\psi_0}$ being the ground state of the pre-quench Hamiltonian, and also by means of a dynamical order parameter. In the first approach, DQPTs manifest as nonanalyticities in the rate function, which are linked to the zeros of the return probability amplitude in the complex time plane~\cite{Heyl2013,Heyl2018,Heyl2019}. In the second approach, a dynamical order parameter~\cite{Sciolla2013,Zhang2017,Bento2024}, often taken as the time-averaged magnetization in a given direction, captures symmetry restoration or breaking induced by the quench. Both perspectives have been extensively explored in reference~\cite{Pappalardi2024}, which shows that, for the anisotropic case, DQPTs occur when the post-quench field crosses the critical value $h_c^d = (J + h_0)/2$.

In the isotropic case, $\gamma = 1$, DQPTs are suppressed for this model. For $h_0 = 0$, the ground state corresponds to one of the degenerate eigenstates of $J_x$. Since $[H, J_x] = 0$ in this limit, the post-quench evolution is trivial: the state only acquires a phase factor while remaining within the same symmetry sector as the initial state. Consequently, the overlap $\bra{\psi_0}\ket{\psi(t)}$ never vanishes, implying the absence of zeros in the return probability amplitude, which is a necessary condition for the occurrence of DQPTs.

\begin{figure}
    \centering
    \includegraphics[width=\linewidth]{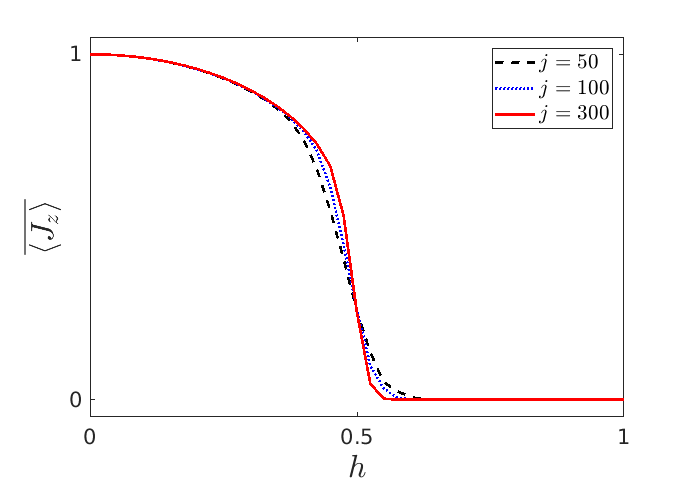}
    \caption{\justifying Dynamical order parameter $\overline{\langle J_z \rangle}$ for the model described by the Hamiltonian~\eqref{H_LMG} as a function of the transverse magnetic field $h$, for quench from $h_0=0$ to $h$, and for different total angular momentum. The convergence to the critical point of the model indicates the critical phenomenon in the thermodynamic limit.}
    \label{OP_LMG}
\end{figure}

To illustrate the occurrence of DQPTs, in Fig.~\ref{OP_LMG} we show the dynamical order parameter for the system described by the Hamiltonian~\eqref{H_LMG}, with $\gamma = 0$, $J = 1$, and for quenches from $h_0 = 0$ to $h$. The dynamical critical point in this case is $h_c^d = 0.5$~\cite{Pappalardi2024}. We observe that the time-averaged magnetization $\overline{\langle J_z \rangle}$ abruptly drops to zero near $h_c^d = 0.5$, confirming the occurrence of a DQPT for the model. This behavior is associated with a dynamical phase transition that drives the system from the dynamical ferromagnetic phase, for $h < 0.5$, where $\overline{\langle J_z \rangle} \neq 0$, to the dynamical paramagnetic phase, where $\overline{\langle J_z \rangle} = 0$~\cite{Bento2024}. 

The $\mathbb{Z}_2$ symmetry is spontaneously broken in the ferromagnetic phase and restored in the paramagnetic phase. In the dynamical context, DQPT can similarly be interpreted as a transition between symmetry-broken and symmetry-restored dynamical phases. As discussed in Ref.~\cite{Bento2024}, the restoration of symmetry in post-quench dynamics is a hallmark of DQPT, reflected in the vanishing of the dynamical order parameter as the system crosses the dynamical critical point.

In order to quantitatively develop these concepts, we consider measures of asymmetry, which quantify how much states, or evolutions, break a given symmetry. With this goal in mind, in the next section we briefly review such measures, originally proposed in Ref.~\cite{Marvian2014}, before applying them to dynamical critical systems. 

\section{Measuring asymmetry}
\label{SecIII}

As discussed in the previous section, symmetries play a very important role in the analysis of DQPTs, as they can be dynamically restored when the system undergoes a transition. Moreover, symmetries in physical systems play a fundamental role in both classical and quantum contexts, especially in relation to conservation laws. According to Noether’s theorem, every continuous symmetry of the dynamics of the system implies the existence of a conserved quantity. Although Noether’s theorem also applies to quantum contexts, it is unable to capture certain essential aspects of quantum evolutions~\cite{Marvian2014}.

A symmetry is specified by a group $G$ whose elements $g \in G$ correspond to the transformations generated by the group. For example, the generators of the rotation group are angular momentum operators $\mathbf{J} = (J_x, J_y, J_z)$. Therefore, a rotation about an axis $\mathbf{n}$ admits a unitary representation given by
\begin{equation}
    U(g) = e^{-i\theta\, \mathbf{J} \cdot \mathbf{n}},
\end{equation}
where $\theta$ is the rotation angle. The action of such a rotation on a quantum state is given by
\begin{equation}
    \rho \;\longrightarrow\; \mathcal{U}_g(\rho) \equiv U(g)\, \rho\, U^\dagger(g).
\end{equation}

In this framework, Noether's conserved quantities, given by the moments of the symmetry generators $L$ of a group $G$, namely $\Tr(\rho L^k)$, do not distinguish between symmetric and asymmetric states, since they are not sensitive to quantum coherence between the eigenstates of the symmetry operator $L$~\cite{Marvian2014}. In other words, two states with the same spectral distribution with respect to $L$ may differ in their symmetric character, depending on the coherence between the eigenstates.

This limitation was the main motivation for the authors of~\cite{Marvian2014} to propose a quantitative framework to assess how much a quantum state breaks a given symmetry, introducing the concept of asymmetry measures as a more complete diagnostic tool.

An asymmetry measure is defined as a function $f: \rho \to \mathbb{R}$ that does not increase under symmetric dynamics~\cite{Marvian2014,Bartlett2006,Gour2008}
\begin{equation}
    f(\rho) \geq f(\Lambda(\rho)),
\end{equation}
where $\Lambda$ is a symmetric map with respect to a group of symmetry $G$. This condition ensures that asymmetry cannot be generated by operations that respect the symmetry considered.

Given this general framework for studying asymmetry, in order to connect it to DQPT, we now need to choose a specific quantifier. Several asymmetry measures have been proposed in the literature~\cite{Gour2008, Gour2009,Vaccaro2008,Skotiniotis2012,Toloui2011}. In this work, we adopt the one presented in~\cite{Marvian2014,Bartlett2006}. Given a quantum state $\rho$ and a generator $L$ of the considered symmetry group, the asymmetry measure is defined by
\begin{equation}
    F_L(\rho) = \|[\rho,L]\|_1.
    \label{FL_function}
\end{equation}
Here, $\|A\|_1 = \text{Tr}\left(\sqrt{A^\dagger A}\right)$ is the $\ell_1$-norm of the operator $A$. Other asymmetry monotones, such as those based on relative entropy or Holevo-type quantities, could also be employed and are expected to display similar qualitative behavior, although the present $\ell_1$-norm measure is particularly convenient because it directly resolves the asymmetry with respect to a chosen generator.

The function $F_L(\rho)$ quantifies how much the state $\rho$ does not commute with the symmetry generator $L$. This lack of symmetry under the action of $G$ manifests itself as coherence in the eigenstates of $L$. If $\rho$ commutes with $L$, then $F_L(\rho) = 0$, the state is symmetric under the transformations generated by $L$. On the other hand, a nonzero value indicates the presence of coherence between different eigenspaces of $L$, signaling that the state breaks the associated symmetry.

\section{Asymmetry and dynamical criticality}
\label{SecIV}

Based on the framework presented in the previous sections, we investigate how the asymmetry measure defined in Eq.~\eqref{FL_function} evolves under the dynamics of the LMG model exhibiting DQPTs. Specifically, we aim to determine whether such an asymmetry measure can serve as an indicator of dynamical quantum phase transitions. In light of the discussion presented in Sec.~\ref{SecII}, we expect a deep connection between asymmetry monotones and dynamical criticality.

As discussed in Sec.~\ref{SecII}, the $\mathbb{Z}_2$ symmetry plays a fundamental role in the identification of DQPTs, with the dynamical restoration of this symmetry constituting a hallmark of the phenomenon. In this sense, focusing on the parity symmetry and on its dynamical evolution can provide insights into the behavior of our model and into the occurrence or absence of a DQPT. We therefore aim to analyze how the parity symmetry manifests itself in the LMG model by means of the quantity defined in Eq.~(\ref{FL_function}).

However, Eq.~(\ref{FL_function}) is defined for generators of Lie groups, which are associated with continuous symmetries~\cite{Marvian2014}. This is not the case for the symmetry of interest here, since it is a discrete symmetry. However, the inversion operation performed by the parity operator can be interpreted as a discrete rotation of angle $\theta=\pi$ around the $x$ axis, represented by a finite element of the rotation subgroup $U(1) \subset SU(2)$ generated by $J_x$.

Given this, we choose to study the generators of the $SU(2)$ group, which describe rotations in spin systems. In collective spin models, these generators are given by the angular momentum operators $J_x$, $J_y$, and $J_z$. This choice allows us to compare the behavior of the asymmetry monotone along different directions in spin space and to assess whether the observed dynamical behavior is common to all rotation generators or if it manifests more prominently for a specific direction. In particular, we investigate whether the most pronounced behavior occurs for $J_x$, in view of its association with parity symmetry. Thus, although the formulation is made in terms of continuous generators, our analysis aims at probing the symmetry associated with the DQPT in the model rather than a genuine rotational symmetry.

We consider two different quenches: from $h_0 = 0$ to $h = 0.2$ (not crossing the critical point) and to $h = 0.8$ (crossing the critical point) and two values of the anisotropy parameter, $\gamma = 0.2$ (strongly anisotropic) and $\gamma = 0.8$ (nearly isotropic). Throughout our analysis, we set $J = 1$ and $j = 100$ in Eq.~(\ref{H_LMG}). These parameter values were chosen to investigate the role of the anisotropy parameter $\gamma$ in two main aspects. First, we ask how the asymmetry measures respond to variations in $\gamma$, particularly as the system approaches the isotropic limit. Secondly, the relationship between the DQPT signatures, which emerge when quenching to the field values $h > h_c^d$, the anisotropic case~\cite{Pappalardi2024}, and the degree of anisotropy.

\begin{figure*}[hbt!]
    \centering
    \includegraphics[width=\linewidth]{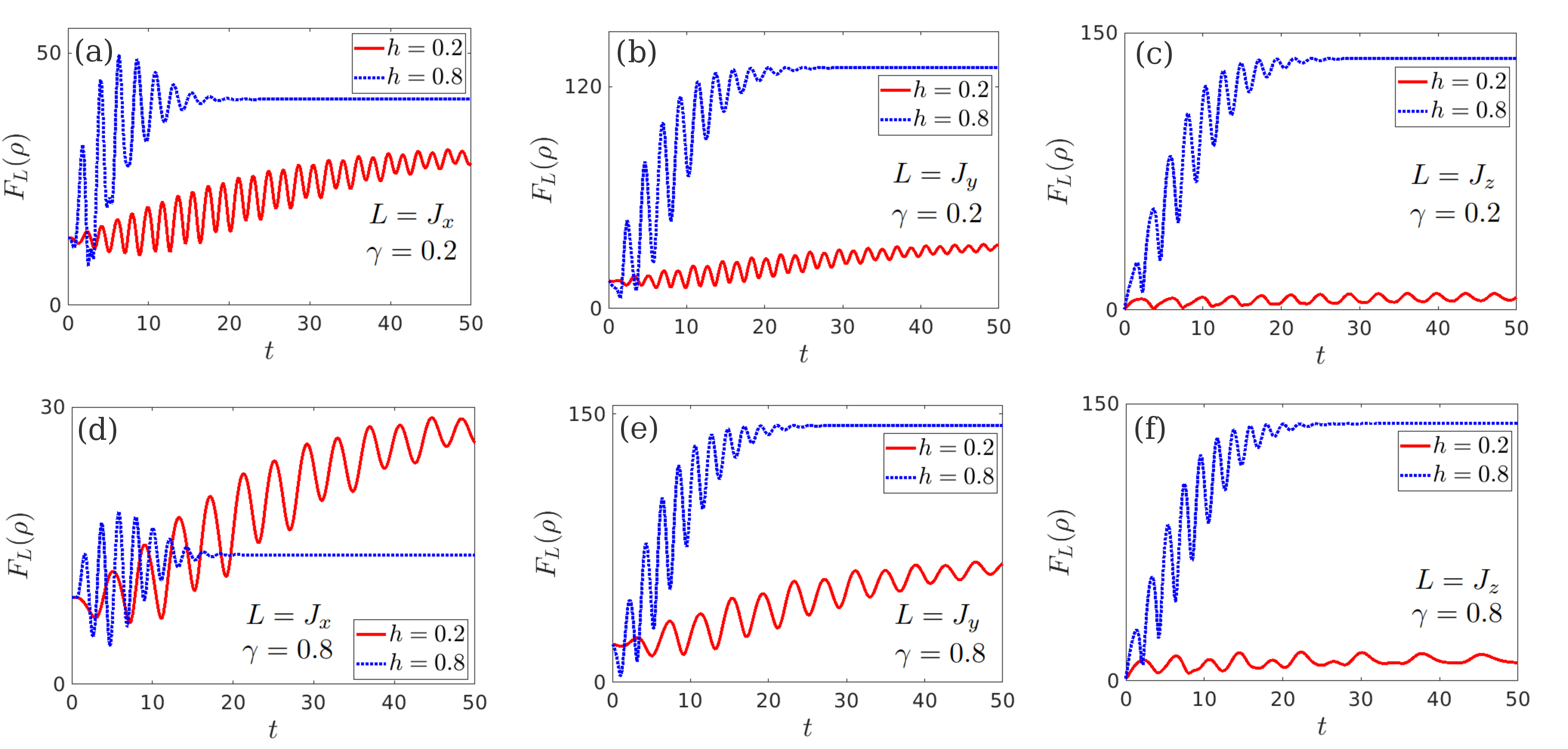}
    \caption{\justifying Time evolution of the asymmetry measure $F_L(\rho)$, Eq.~(\ref{FL_function}), for $L = J_x$, $J_y$, and $J_z$ shown in the first, second, and third columns, respectively, and for two different anisotropy parameters: $\gamma = 0.2$ (top row) and $\gamma = 0.8$ (bottom row). In all plots, two quenches are considered: $h = 0.2$ (red solid line) and $h = 0.8$ (blue dotted line).}
    \label{FL_plots}
\end{figure*}

We begin by analyzing the behavior of the asymmetry measure $F_L(\rho)$ as a function of the intensity of the quench, the anisotropy parameter $\gamma$, and the choice of generator $L$. As a main feature, in Fig.~\ref{FL_plots}, we observe that the oscillations of $F_L(\rho)$ tend to be damped more rapidly when the quench crosses the dynamical critical point $h_c^d$. This behavior is observed in different values of $\gamma$ and for all three generators considered. In other words, the dynamics saturates much faster when we cross the critical point. 

For the cases where $L = J_y$ and $L = J_z$, the behavior of the asymmetry measure is qualitatively similar. In both directions, and for both $\gamma = 0.2$ and $\gamma = 0.8$, we observe rapid growth in $F_L(\rho)$ when quenching is performed in $h > h_c^d$. After this initial increase, the measure stabilizes at a plateau. This behavior is similar to previous results reported in~\cite{Nascimento2024}, where it was shown that DQPTs induce a fast stabilization of the dynamics of the system, reflected in the saturation of the upper bound of entropy production. Our findings indicate that the asymmetry along the $y$ and $z$ directions follows the same pattern: the system becomes more asymmetric immediately after the quench but quickly reaches a steady regime. However, for quenches that do not cross the critical point, the rate of growth in asymmetry is much slower, and the system tends to remain closer to the initial asymmetry. This is a direct consequence of the change in symmetry when we cross the critical point, pointing to the phase transition.

A different scenario emerges when considering $L = J_x$, which is directly associated with the transverse field term in the Hamiltonian~\eqref{H_LMG}. In this case, two key differences are observed. First, the anisotropy parameter plays a more prominent role. For $\gamma = 0.2$, the behavior resembles that seen for $J_y$ and $J_z$: the asymmetry increases significantly when $h > h_c^d$. However, for $\gamma = 0.8$, this pattern is inverted and larger asymmetry values are observed when the quench does not cross the critical point. Second, the quantitative difference between the two quenches considered is less pronounced for $J_x$ compared to the other generators, regardless of the anisotropy value.

In summary, Fig.~\ref{FL_plots} shows the asymmetry measures associated with rotations around the three Cartesian axes; that is, considering the generators of the rotation group, $J_x$, $J_y$, and $J_z$, they are sensitive to the critical quench by exhibiting significant changes in their behavior. It should be noted that, for the case with the highest anisotropy, $\gamma = 0.2$, the behavior of the three generators was similar, with a rapid growth of $F_L(\rho)$ for the quench $h=0.8 > h_c^d$, in contrast to the small growth observed for the quench $h=0.2 < h_c^d$. This indicates that crossing the dynamical critical point induces a fast increase in the coherence of $\rho$ in the eigenbasis of the generator $L$, leading to an enhancement of asymmetry in the model.

An exception occurs in the case shown in Fig.~\ref{FL_plots}(d), which presents the asymmetry measure with respect to the generator $L=J_x$ for $\gamma = 0.8$. This is the only situation in which the critical quench does not produce more coherence in the state. We attribute this to the discussion in Sect.~\ref{SecII} about the possibility of rewriting Eq.~\eqref{H_LMG} solely in terms of $J_x$ in the isotropic case. In this regime, near the isotropic model, the ground state exhibits lower asymmetry with respect to $L=J_x$, and a stronger quench that increases the contribution of $J_x$ in the Hamiltonian tends to preserve the system's symmetry. From the perspective of symmetry dynamics, this can be interpreted as a situation where evolution does not effectively populate states with different $J_x$ eigenvalues, preventing the generation of coherence in that basis and, consequently, limiting the growth of asymmetry.

Given that we are dealing with a non-equilibrium quantum system undergoing a sudden quench, it is natural to consider the time average of the quantity under investigation. As the temporal evolution of such quantities typically exhibits pronounced oscillations, the time-averaged value provides a more reliable indicator of the long-term behavior of the system.

Following this reasoning, we compute the time average of the asymmetry measure $F_L(\rho)$ for the cases discussed in Fig.~\ref{FL_plots}. This average is defined as
\begin{equation}
    \overline{F_L(\rho)} = \frac{1}{T} \int_{0}^{T} dt  F_L(\rho).
    \label{TAF_function}
\end{equation}

\begin{figure*}
    \centering
    \includegraphics[width=\linewidth]{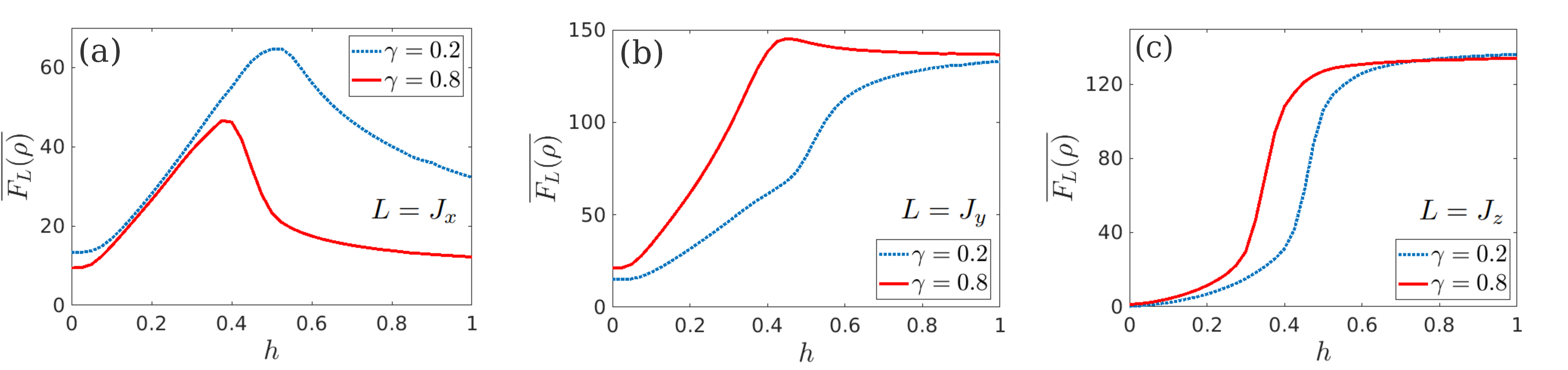}
    \caption{\justifying Time average of the asymmetry measure $\overline{F_L(\rho)}$ in terms of quench parameter for the three generators,  $(a) \,\, L=J_x$, $(b) \,\, L=J_y$ and $(c)\,\,L=J_z$, of the rotation group and two anisotropy parameters: $\gamma = 0.2$ (blue dotted line) and $\gamma = 0.8$ (red solid line).}
    \label{TAFL_plots}
\end{figure*}

The results of Eq.~(\ref{TAF_function}), for $L=J_x$, $J_y$, and $J_z$, are shown in Fig.~\ref{TAFL_plots}. Two important features can be observed in these graphs: first, there is a noticeable change in the behavior of $\overline{F_L(\rho)}$ as the quench parameter $h$ increases; and second, the shift of the point at which this change occurs as a function of the anisotropy parameter $\gamma$. These behaviors reveal additional insight into the dynamics of the system.

In Fig.~\ref{TAFL_plots}(c), we observe that $\overline{F_{J_z}(\rho)}$ remains small for quenches to low values of the transverse field $h$, regardless of the anisotropy parameter $\gamma$. However, as $h$ increases, the asymmetry measure shows a rapid increase around $h \approx 0.5$ for $\gamma = 0.2$, and $h \approx 0.4$ for $\gamma = 0.8$. For higher values of $h$, the curves tend to stabilize, indicating a steady regime in the long-time dynamics.

A similar pattern is found for the generator $J_y$, as shown in Fig.~\ref{TAFL_plots}(b). In this case, the time-averaged asymmetry also increases with $h$, particularly around the same critical regions as for $J_z$. The main distinction appears for $\gamma = 0.8$, where $\overline{F_{J_y}(\rho)}$ exhibits a slight decrease after reaching a peak near $h = 0.4$.

The behavior of $\overline{F_{J_x}(\rho)}$, shown in Fig.~\ref{TAFL_plots}(a), is especially interesting due to the direct connection between $J_x$ and the transverse field term in the Hamiltonian~\eqref{H_LMG}. Unlike the previous cases, the lowest values of $\overline{F_{J_x}(\rho)}$ occur for $\gamma = 0.8$, that is, when the system is closer to the isotropic limit. This difference can be associated with the fact that in this regime the Hamiltonian approach $J_x$-symmetry, as mentioned previously. Thus, increasing the transverse field term $hJ_x$ contributes less to the generation of asymmetry. Despite this difference, the location of the turning point in the asymmetry measure remains close to those observed for $J_y$ and $J_z$. For $\gamma = 0.8$, the maximum is found around $h \approx 0.4$, followed by a decrease as $h$ increases further. A similar trend is seen for $\gamma = 0.2$, with the peak occurring near $h \approx 0.5$.

An important question that arises from the previous analysis concerns the physical meaning of the observed change in the behavior of the time-averaged asymmetry measure. In particular, why does this change occur at different values of $h$ depending on the anisotropy parameter $\gamma$?

The behavior of the time-averaged lower bound of entropy production, $\overline{\langle \Sigma \rangle}$, in the LMG model under similar quench conditions was investigated in Ref.~\cite{Nascimento2024}. Entropy production is quantified through a geometric lower bound, in terms of the Bures angle between the time-evolved state $\rho$ and the corresponding equilibrium reference state $\rho^{eq}$
\begin{equation}
    \langle \Sigma \rangle \geq s\left( \frac{2}{\pi} \mathcal{L}(\rho,\rho^{eq}) \right)
\end{equation}
where $\mathcal{L}(\rho,\rho^{eq})$ is the Bures angle and $s(x) = \min_{x<r<1} S((r-x,1-r+x)\vert\vert (r,1-r))$~\cite{Deffner2013}.
The peak in $\overline{\langle \Sigma \rangle}$ was shown to be associated with the occurrence of a dynamical quantum phase transition, the maximum located near $h = 0.5$, which is the dynamical critical point of the model under the conditions considered. However, this analysis was performed for a single fixed anisotropy parameter value, $\gamma = 0.5$.

\begin{figure}[h]
    \centering
    \includegraphics[width=\linewidth]{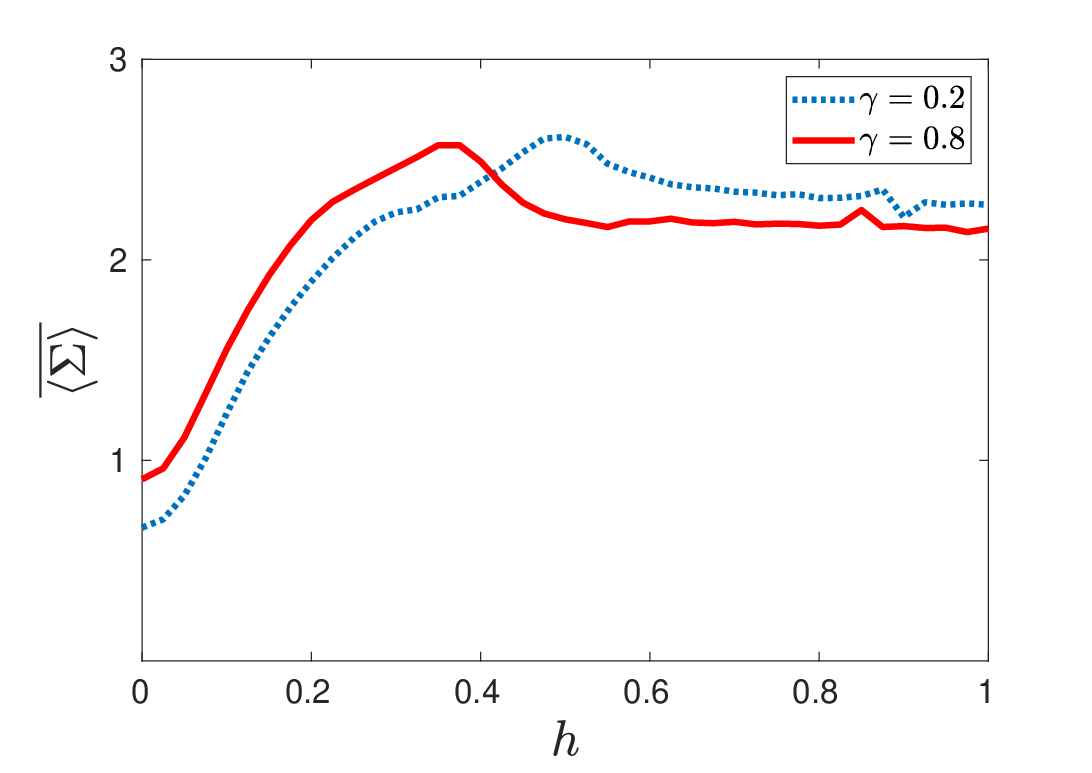}
    \caption{\justifying Time average of Entropy Production $\overline{\langle\Sigma\rangle}$, according reference~\cite{Nascimento2024}, for $j=100$ and two different anisotropy parameters: $\gamma=0.2$ (blue dotted line) and $\gamma=0.8$ (red solid line).}
    \label{TAEPh}
\end{figure}

Based on this result, we now explore whether the same correspondence holds for different values of the anisotropy parameter. Specifically, we analyze the behavior of $\overline{\langle \Sigma \rangle}$ for $\gamma = 0.2$ and $\gamma = 0.8$, using the same quench protocols applied to the analysis of asymmetry. The results are presented in Fig.~\ref{TAEPh}. As shown, the location of the peak in $\overline{\langle \Sigma \rangle}$ changes as the anisotropy parameter varies. Interestingly, this shift closely mirrors the behavior observed in the asymmetry measures $\overline{F_L(\rho)}$, with both quantities exhibiting maxima at similar values of $h$.

\begin{figure}[h]
    \centering
    \includegraphics[width=\linewidth]{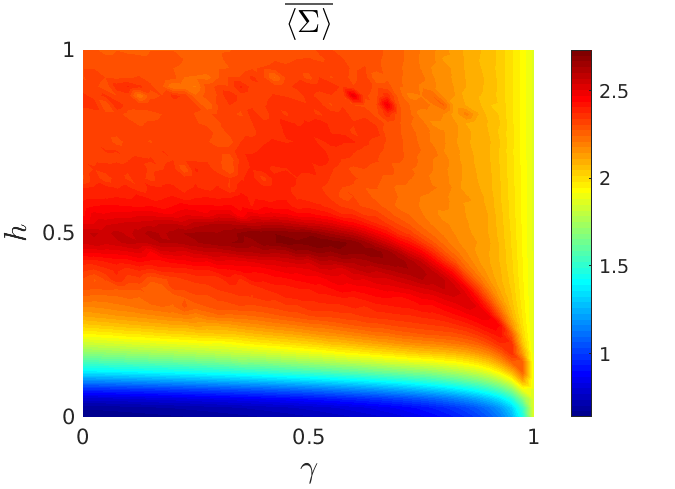}
    \caption{\justifying Time average of Entropy Production $\overline{\langle\Sigma\rangle}$ in function of $\gamma$ and $h$, for $j=100$. We vary $\gamma \in (0,1)$ and $h \in (0,1)$.}
    \label{TAEPhg}
\end{figure}

\begin{figure*}[hbt!]
    \centering
    \includegraphics[width=\linewidth]{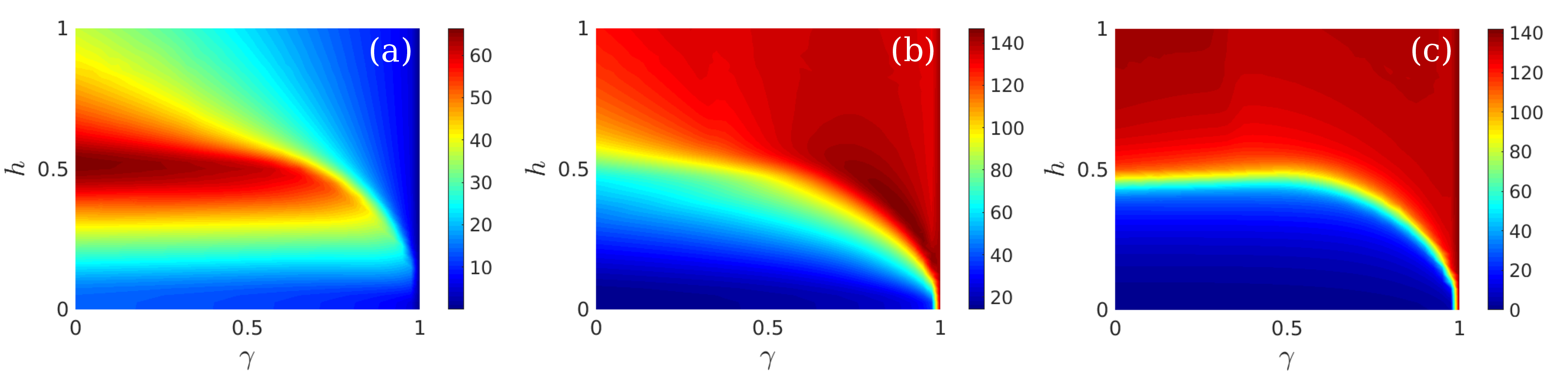}
    \caption{\justifying Time average of the asymmetry measure $\overline{F_L(\rho)}$ in terms of quench parameter $h$ and anisotropy parameter $\gamma$, for the three generators,  $(a) \,\, L=J_x$, $(b) \,\, L=J_y$ and $(c)\,\,L=J_z$.  We vary $\gamma \in (0,1)$ and $h \in (0,1)$ and fixed $j=100$.}
    \label{Fhg_plots}
\end{figure*}

To investigate the behavior of $\overline{\langle \Sigma \rangle}$ in a more comprehensive way and the resulting change in the dynamical critical point of the model, we plot the time average of $\langle \Sigma \rangle$ over the range $\gamma \in (0,1)$, which spans from the maximally anisotropic model to the isotropic limit, and $h \in (0,1)$, which includes the critical quench discussed in Ref.~\cite{Nascimento2024}. The result, shown in Fig.~\ref{TAEPhg}, clearly reveals a shift of the peak position as a function of $\gamma$, particularly for $\gamma > 0.5$, with the peak progressively disappearing as $\gamma \to 1$, i.e., in the isotropic limit. This suggests that the anisotropy parameter plays an important role in modifying the location of the dynamical critical point $h_c^d$.

In view of the behavior presented in Fig.~\ref{TAEPhg} and our interest in assessing whether the asymmetry measure can serve as an indicator of DQPT, we compute the time-averaged asymmetry measure $\overline{F_L(\rho)}$ for the generators $J_x$, $J_y$, and $J_z$, varying both the anisotropy parameter $\gamma$ and the transverse field $h$ over the same range.

Figure~\ref{Fhg_plots} provides this global view of the time-averaged asymmetry measure $\overline{F_L(\rho)}$ in the $(h,\gamma)$ plane. A common feature shared by all cases is the presence of a clear change in the behavior of $\overline{F_L(\rho)}$ as the transverse field $h$ is varied, whose location depends on the anisotropy parameter $\gamma$. For all generators, this change is progressively shifted as $\gamma$ increases and tends to disappear as the system approaches the isotropic limit $\gamma \to 1$. This behavior is consistent with the trends observed previously in the time-averaged entropy production, Fig.~\ref{TAEPhg}, and reflects the strong influence of anisotropy on the dynamical critical properties of the model.

The case $L=J_x$, shown in Fig.~\ref{Fhg_plots}(a), deserves special attention due to its direct connection with the $\mathbb{Z}_2$ parity symmetry in our model. Although the signatures in $\overline{F_{J_x}(\rho)}$ are less pronounced than those observed for other generators, the changes in their behavior occur in the same parameter regions where dynamical criticality is expected. In particular, the sharp peak of $\overline{F_{J_x}(\rho)}$ coincides with the parameter region associated with the dynamical restoration of the parity symmetry, that is, with the region where the DQPT takes place, as independently identified in Fig.~\ref{TAEPhg}. Moreover, after the peak, the asymmetry no longer grows significantly, reflecting that the system ceases to generate parity-related asymmetry at the same rate. This behavior contrasts with the cases $L=J_y$ and $L=J_z$, for which no symmetry restoration is expected, and where the observed changes in $\overline{F_L(\rho)}$ simply signal a modification of the dynamical regime rather than the restoration of an underlying symmetry.

\begin{figure}[h]
    \centering
    \includegraphics[width=\linewidth]{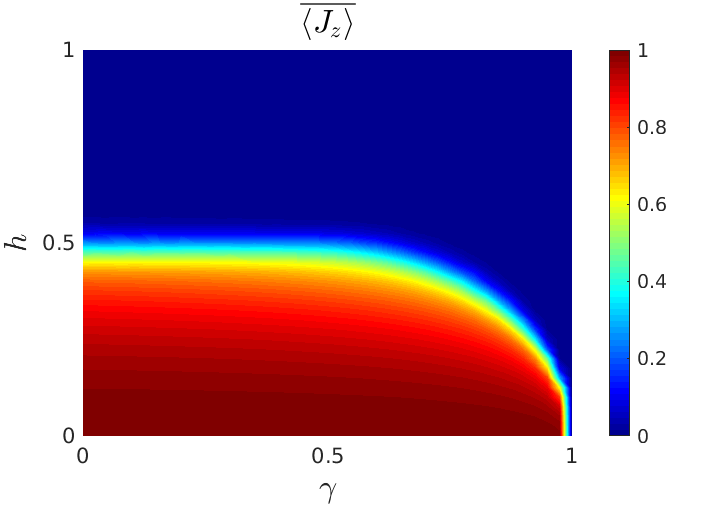}
    \caption{\justifying Dynamic order parameter $\overline{\langle J_z\rangle}$ in terms of quench parameter $h$ and anisotropy parameter $\gamma$. We use the same range that in Figs.~\ref{TAEPhg} and~\ref{Fhg_plots}: $\gamma \in (0,1)$ and $h \in (0,1)$ and fixed $j=100$.}
    \label{OrderPar_hg}
\end{figure}

In contrast, the asymmetry measure associated with $L=J_z$, Fig.~\ref{Fhg_plots}(c), exhibits the most pronounced and well-defined transition line in the $(h,\gamma)$ plane. This behavior closely mirrors that of the dynamical order parameter, Fig.~\ref{OrderPar_hg}, which is also defined in terms of the collective magnetization along the $z$ direction and characterizes the DQPT through its decay to zero. As a result, $\overline{F_{J_z}(\rho)}$ provides a highly sensitive diagnostic of the dynamical transition. Nevertheless, this enhanced sensitivity should be understood as a consequence of the close relationship between $J_z$ and the order
parameter, rather than as an indication that rotations around the $z$ axis define the symmetry underlying the DQPT.

Finally, we use the dynamical order parameter as an independent reference to validate the critical nature of the region identified through the asymmetry measure. Importantly, the critical lines defined by the decay of the order parameter coincide with the parameter region where the distinctive behavior of $\overline{F_{J_x}(\rho)}$ is observed. This agreement confirms that the region identified through the asymmetry measure associated with $J_x$ corresponds to a regime consistent with dynamical critical behavior, thus validating the interpretation of $\overline{F_{J_x}(\rho)}$ as a signature of the dynamical restoration of parity symmetry.

These consistent results from three independent quantities: entropy production, asymmetry, and dynamical order parameter demonstrate that the time-averaged asymmetry measure $\overline{F_L(\rho)}$ captures the essential features of dynamical quantum criticality. This reinforces its potential role as a robust indicator of DQPT in the LMG model.

Finally, we note that the dynamical critical value $h_c^{\mathrm d}$ is closely related to the excited-state quantum phase transition (ESQPT) of the LMG model~\cite{Corps2022}. In the fully anisotropic case, this connection can be made explicit: the dynamical critical point is reached when the energy of the initial state, evaluated with the post-quench Hamiltonian, coincides with the ESQPT separatrix energy. Equating the post-quench energy density with the separatrix energy in the semiclassical limit yields
\[
h_c^{\mathrm d}=\frac{J+h_0}{2},
\]
demonstrating that the critical value observed in the dynamics exactly corresponds to the crossing of the ESQPT energy. As the anisotropy parameter $\gamma$ is varied, the location of the ESQPT singularity changes, leading to the shift of the dynamical critical region observed in Figs.~5--7. In the isotropic limit $\gamma \to 1$, the ESQPT disappears together with the dynamical critical behavior.

\section{Conclusion}
\label{SecV}

The dynamical restoration and breaking of symmetries are central features of dynamical quantum phase transitions, yet their quantitative characterization remains a challenge. In this work, we have shown that asymmetry measures provide a natural and physically transparent framework to probe these phenomena, establishing a direct connection between symmetry, quantum coherence, and nonequilibrium critical dynamics in the Lipkin-Meshkov-Glick model.

By analyzing asymmetry measures associated with rotations around the three Cartesian axes, we demonstrated that different generators capture complementary aspects of the dynamical critical behavior. This approach allowed us to disentangle the role of the parity symmetry $\mathbb{Z}_2$, associated with the generator $J_x$, from that of observables such as $J_z$, which are more directly related to the dynamical order parameter and therefore exhibit sharper numerical signatures of the transition. This distinction highlights that sensitivity to criticality and sensitivity to symmetry restoration are, in general, not equivalent notions.

Our results further show that both the quench strength and the anisotropy parameter $\gamma$ play a decisive role in shaping the dynamical response of the system. In particular, increasing anisotropy shifts the location of the dynamical critical region and ultimately suppresses the dynamical criticality as the isotropic limit is approached. This behavior underscores the importance of anisotropy as a control parameter for dynamical phase transitions in collective spin models with long-range interactions.

Analysis of time-averaged asymmetry measures revealed abrupt changes in their behaviour in the dynamical critical region, in close agreement with independent signatures provided by entropy production~\cite{Nascimento2024}. This correspondence supports a thermodynamic interpretation of the generation of asymmetry, suggesting that DQPTs are accompanied by enhanced irreversibility and rapid redistribution of coherence across symmetry sectors. Although the asymmetry associated with $J_z$ provides the most pronounced signal due to its close connection with the dynamical order parameter, this enhanced sensitivity should be understood as a diagnostic feature rather than as an indication of the symmetry governing the transition itself.

In contrast, the asymmetry associated with $J_x$ plays a conceptually distinct role. Although it does not maximize sensitivity to the transition, its behavior directly reflects the dynamical restoration of the $\mathbb{Z}_2$ parity symmetry. The emergence of a characteristic peak followed by saturation in the time-averaged asymmetry $\overline{F_{J_x}(\rho)}$ coincides with the critical region identified by the dynamical order parameter, confirming that this quantity captures the symmetry-driven nature of the transition. This demonstrates that asymmetry measures not only act as indicators of dynamical quantum phase transitions but also provide insight into the specific symmetry mechanisms underlying critical dynamics.

An important aspect of the present approach is that its applicability is not restricted to the LMG model, but rather depends on the presence of relevant symmetry structures in the dynamical evolution. In this sense, asymmetry measures are expected to provide useful information in other systems exhibiting dynamical criticality whenever the transition is accompanied by a redistribution of the state across symmetry sectors, associated to an appropriate generator. On the other hand, in models with finite-range interactions or non-integrable dynamics, where the role of symmetry may be less direct or even absent, the corresponding signatures may not be present. These observations indicate that the usefulness of asymmetry as a diagnostic tool is not universal, but instead depends on the extent to which symmetry-related mechanisms underlie the critical dynamics.

Although in the present work we focus on sudden quenches, it is natural to ask how the asymmetry behaves under finite-time ramps. In the adiabatic limit, the evolution remains within the instantaneous symmetry sector of the initial state and the characteristic signatures associated with the restoration of parity symmetry are expected to disappear. For ramps of finite duration, however, nonadiabatic excitations generated near the critical region should produce an excess asymmetry relative to the adiabatic state. Therefore, asymmetry measures may also provide a probe of the breakdown of adiabaticity and of the finiteness of the driving protocol.

Several open questions naturally arise from this work. An important direction is the extension of our analysis to alternative asymmetry measures, particularly those based on Holevo-type quantities, which have been shown to possess a direct thermodynamic interpretation in terms of entropy~\cite{Ferrari2025}. More broadly, it would be interesting to investigate the role of asymmetry in mixed-state and finite-temperature DQPTs, as well as in open quantum systems where dissipation and decoherence compete with symmetry-driven dynamics. It is also interesting to observe that asymmetry measures based on generalized relative entropies have been linked to quantum speed limits and discussed in the context of thermodynamic behavior in unitary processes~\cite{Pires2021}. This suggests that further study of such measures may deepen our understanding of the interplay between symmetry breaking, nonequilibrium thermodynamics, and dynamical criticality in quantum many-body systems. Finally, exploring the universality of asymmetry-based signatures across different models and symmetry classes, as well as their experimental accessibility in platforms such as cold atoms and trapped ions, remains an exciting avenue for future research.

\begin{acknowledgments}
This research was funded by CNPq through grant 308065/2022-0, the National Institute of Science and Technology for Applied Quantum Computing through CNPq grant 408884/2024-0, the Coordination of Superior Level Staff Improvement (CAPES), and FAPEG through grant 202510267001843.
\end{acknowledgments}


\end{document}